\documentclass[prd, twocolumn, nofootinbib]{revtex4}
\usepackage{graphicx}
\usepackage{dcolumn}
\usepackage{epsfig} 
\usepackage{amsmath}
\usepackage{amsfonts}
\usepackage{graphicx}
\usepackage{amssymb}

\begin{document}

\title{Excursion Set Halo Mass Function and Bias in a Stochastic
  Barrier Model of Ellipsoidal Collapse}
\author{P.S. Corasaniti}
\author{I. Achitouv}
\address{Laboratoire Univers et Th\'eories (LUTh), UMR 8102
  CNRS, Observatoire de Paris, Universit\'e Paris Diderot, \\ 5 Place Jules Janssen, 92190 Meudon, France}
\begin{abstract}
We use the Excursion Set formalism to compute the properties of the halo mass distribution
for a stochastic barrier model which encapsulates the main 
features of the ellipsoidal collapse of dark matter halos. 
Non-markovian corrections due to the sharp filtering of the linear density field in
real space are computed with the path-integral technique introduced
by Maggiore \& Riotto \cite{MaggioreRiotto2010a}. Here, we provide a
detailed derivation of the results presented in \cite{CorasanitiAchitouv2010} and extend the mass
function analysis to higher redshift. We also derive
an analytical expression for the linear halo bias. We find the
analytically derived mass function to be in 
remarkable agreement with N-body simulation data from Tinker et al. 
\cite{Tinker2008} with differences $\lesssim
5\%$ over the range of mass probed by the simulations. The excursion
set solution from Monte Carlo generated random walks shows the same
level of agreement, thus confirming the validity of the
path-integral approach for the barrier model considered here.
Similarly the analysis of the linear halo bias shows deviations no greater than $20\%$. 
Overall these results indicate that 
the Excursion Set formalism in combination with a realistic modeling
of the conditions of halo collapse can provide
an accurate description of the halo mass distribution.
\end{abstract} 

\maketitle

\section{Introduction}
Dark matter (DM) is an essential ingredient of the Standard
Model of Cosmology. Observations provide strong evidence that about 
$90\%$ of the matter in cosmic structures consists of such an
invisible component \cite{Spergel2003,Tegmark2004,Clowe2006,Massey2007}. 
Central to the DM paradigm is the idea that 
initial DM density fluctuations grow under gravitational instability
fostering the collapse of baryonic matter.
At late time, the gravitational infall becomes 
highly non-linear and DM particles violently relax into virialized objects, the halos.
These are the building blocks in which cooling baryonic gas
falls in to form the stars and galaxies that we observe today. 
Hence, the study of the halo mass distribution is 
of primary importance in Cosmology and an accurate modeling of the
halo mass function has become essential to disclosing the complex 
mechanisms of the cosmic structure formation
as well as probing the physics of the early universe.
For example, recent studies have focused on the computation
of the mass function for non-Gaussian initial
conditions which are a direct signature of inflationary physics
(see e.g. \cite{Loverde2008,MaggioreRiotto2010c,Damico2011}). 

The upcoming generation of galaxy cluster surveys 
(see e.g. \cite{Pierre2011,ACT2010SPT2010}) will directly probe
the distribution of massive halos, thus providing
cosmological constraints complementary to those inferred from
measurements of the Cosmic Microwave Background (CMB) radiation and cosmic 
distance indicators. 

The properties of the DM halo mass distribution have been mainly investigated 
using high-resolution numerical N-body simulations. 
These studies have determined the halo
mass function to a few percent accuracy level and provided insights
on its redshift and cosmology dependence
\cite{Tinker2008,Courtin2010,Crocce2010,Suman2011}. 
In contrast, the development of purely theoretical studies has
lagged behind. To date, a robust theoretical description of the halo
mass function is still missing. For instance, we do not have a complete 
understanding of the relation between the conditions that lead 
to the formation of halos and the form of the mass function 
as obtained from N-body simulations. The numerical analysis 
usually limits to providing fitting formulae which depends 
on several ad-hoc parameters. 

The seminal work by Press \& Schechter (PS) \cite{PressSchechter1974} is the first to attempt a
derivation of the halo mass function from the statistical properties of the
initial dark matter density fluctuation field. The idea is that halos form
in regions in which the linearly extrapolated density field, smoothed on 
a given scale, lies above a given critical threshold of collapse, such
as that predicted from the spherical collapse model \cite{GunnGott1972}. 
The PS formula explicitly depends on such a threshold, which introduces
an exponential cut-off in the high-mass end as confirmed by N-body
simulation analysis. However, the PS computation suffers of an
inconsistent behavior when the variance becomes arbitrarily large. 
In this asymptotic regime it is natural to expect that the mass
fraction tends to $1$ (i.e. the higher the amplitude of the density
fluctuations the larger is the fraction of mass in halos). 
However, in the PS case one finds $1/2$, thus suggesting that half 
of the mass in halos has been somehow miscounted.
This is the so called `cloud-in-cloud' problem 
(see e.g. \cite{PeacockHeavens1990} and discussion therein). 

The formulation of the Excursion Set theory by Bond et
al. \cite{Bond1991} has provided the Press-Schecther approach with 
a powerful mathematical formalism in which the computation of the
mass function is reduced to solving a stochastic calculus problem
(for an exhaustive review of the formalism see \cite{Zentner2007}). 
As shown in \cite{Bond1991} the smoothed density fluctuation field
behaves as a stochastic variable performing 
a random walk as function of the smoothing scale. Then, 
halos are associated to random trajectories which 
first cross a critical density threshold of collapse. It is the
requirement of first-crossing which provides the solution to
the `cloud-in-cloud' problem. Nevertheless, 
the computation of the halo mass function for realistic halo
mass definitions has remained a challenging task. In fact, 
in the Excursion Set theory the mass of a halo depends on the form of
the filter function, with the latter determining the behavior of the
random walks. For realistic mass definitions the associated filters 
cause the random walks to be correlated, consequently the mass function
can be estimated only through numerical Monte Carlo simulations
\cite{Bond1991,Percival2001}. Because of this, a thorough systematic
comparison of the Excursion Set mass function obtained
accounting for such correlations against N-body simulation data 
has never being performed.

A major breakthrough in this direction has been recently made by Maggiore \& Riotto \cite{MaggioreRiotto2010a} who have 
introduced path-integral techniques to perform an analytic computation
of the halo mass function for generic filters. This allows us to 
consistently compare the Excursion Set model predictions with N-body
data as well with astrophysical measurements of halo abundances. 

Here, we compute the halo mass function and the linear halo bias for a
barrier model which captures the main features of the ellipsoidal
collapse of halos over a large range of masses. In this paper we
also provide a detailed derivation of the results presented in \cite{CorasanitiAchitouv2010}.

The paper is organized as follows. In Section~\ref{excform} we briefly
introduce the Excursion Set formalism, in Section~\ref{barrmod} we
discuss the modeling of the non-spherical collapse of halos and the
computation of the mass function. In Section~\ref{nonmark} we present the calculation of
the corrections due to the filter function. In Section~\ref{numan} we
discuss the results of the comparison with N-body simulation data.
In Section~\ref{secbias} we present the computation of the halo
bias. Finally, we present our conclusion in Section~\ref{conclu}.

\section{The Halo Mass Function and The Excursion Set Formalism}\label{excform}
In the Press-Schecther approach halos form from regions of the
smoothed linear density fluctuation field which lie above a given
density threshold. In such a case, the number of halos in the mass range 
$[M,M+dM]$ can be written as
\begin{equation}
\frac{dn}{dM}=f(\sigma)\frac{\bar{\rho}}{M^2}\frac{d\log{\sigma^{-1}}}{d\log{M}},\label{mf}
\end{equation}
where $\bar{\rho}$ is the background matter density and $\sigma(M)$ is the root-mean-square
fluctuation of the linear dark matter density field smoothed on a
scale $R(M)$ (containing a mass $M$) which is given by
\begin{equation}
\sigma^2(M)\equiv S(M)=\frac{1}{2\pi^2}\int dk~k^2P(k)\tilde{W}^2[k,R(M)],
\end{equation}
where $P(k)$ is the linear matter power spectrum at redshift $z=0$ and $\tilde{W}(k,R)$ is
the Fourier transform of the smoothing (filter) function
in real space. The function $f(\sigma)$ in Eq.~(\ref{mf}) is usually
dubbed as `multiplicity function' and encodes all the effects responsible for the
formation of halos. It is given by
$f(\sigma)=2\sigma^2\mathcal{F}(\sigma^2)$ 
where $\mathcal{F}(S)=dF/dS$ and $F(S)$ gives the fraction of mass 
elements in halos with mass $>M$. The goal of the Excursion Set 
computation is to evaluate $\mathcal{F}(S)$ and infer $f(\sigma)$.
Hereafter, we will refer to $f(\sigma)$ simply as the mass function. 

Let us briefly review the Excursion Set formalism. 
The density perturbation is defined as
$\delta(\textbf{x})=[\rho(\textbf{x})-\bar{\rho}]/\bar{\rho}$, where
$\rho(\textbf{x)}$ is the local density at the comoving position
$\textbf{x}$. Then, the smoothed density fluctuation field on a scale $R$ is
then given by
\begin{equation}
\delta(\textbf{x},R)=\int d^3y\,W(|\textbf{x-y}|,R)\delta(\textbf{y}),\label{fs}
\end{equation}
where $W(x,R)$ is the filter function in real space. In Fourier space,
Eq.~(\ref{fs}) reads as
\begin{equation}
\delta(\textbf{x},R)=\frac{1}{(2\pi)^3}\int d^3k\,\tilde{W}(k,R)\tilde{\delta}(\textbf{k}) e^{-i\textbf{k x}}
\end{equation}
where $\tilde{\delta}(\textbf{k})$ is the Fourier transform of
$\delta(\textbf{x})$. Bond et al. \cite{Bond1991} showed that by taking the
derivative with respect to $R$, at any
point in space  $\delta(\textbf{x},R)$ obeys a Langevin equation of the form:
\begin{equation}
\frac{\partial\delta}{\partial R} =\zeta(R),  \label{langevin_zeta}
\end{equation}
where
\begin{equation}
\zeta(R)\equiv \frac{1}{(2\pi)^3}\int d^3k\, \tilde{\delta}(\textbf{k})\frac{\partial \tilde{W}}{\partial R}e^{-i\textbf{k x}},
\end{equation}
is a noise term, whose properties depends on the statistics of the
underlying density field. For initial Gaussian density fluctuations,
\begin{equation}
\begin{array}{lr}
\langle\tilde{\delta}(\textbf{k})\rangle=0\,\,\mbox{and} &
\langle\tilde{\delta}(\textbf{k})\tilde{\delta}(\textbf{q})\rangle=(2\pi)^{3}\delta_{D}(\textbf{k}+\textbf{q})P(k),\nonumber
\end{array}
\end{equation}
where $\delta_D$ is the Dirac-function, thus implying that $\langle\zeta(R)\rangle=0$ and
\begin{equation}
\langle \zeta(R_{1}) \zeta(R_{2}) \rangle=\frac{1}{2\pi^2}\int dk\,
k^2 P(k)\frac{\partial \tilde{W}}{\partial R_{1}}  \frac{\partial
  \tilde{W}}{\partial R_{2}}.\label{noisec}
\end{equation}

Halos corresponds to those random walks which hit for the first time an 
absorbing barrier whose value is specified by a critical density threshold of
collapse. In the spherical collapse model this is usually denoted by $\delta_c$. 

From Eq.~(\ref{noisec}) it is evident
that the nature of the random walks depends on the form of 
the filter function. For the time being, let us assume the sharp-k filter,
\begin{equation}
\tilde{W}(k,R)=\theta(1/R-k),
\end{equation}
then by substituting in Eq.~(\ref{noisec}) and after some algebric manipulation
we obtain
\begin{equation}
\frac{\partial\delta}{\partial S} =\eta_\delta(S),\label{langevin_eta}
\end{equation}
with $\langle\eta_\delta(S)\rangle=0$ and
$\langle\eta_\delta(S_1)\eta_\delta(S_2)\rangle=\delta_D(S_1-S_2)$. Hence,
for the sharp-k filter $\eta(S)$ is a
white noise and $\delta$ performs a simple Markov random walk
as function of the variance $S$. 

Let us define $\Pi$ the probability of a trajectory to have value
$\delta$ at time $S$. Then, the probability distribution associated 
to trajectories obeying Eq.~(\ref{langevin_eta}), which start at $\delta(0)=0$\footnote{At
  very large scales $S\rightarrow 0$ and density perturbation
  $\delta\rightarrow 0$.} and are absorbed by the spherical collapse barrier
$B=\delta_c$, is given by the Fokker-Planck
\begin{equation}
\frac{\partial\Pi}{\partial S}=\frac{1}{2}\frac{\partial^2\Pi}{\partial\delta^2},\label{FokkerPlanck}
\end{equation}
with initial
condition $\Pi(\delta,S=0)=\delta_{D}(\delta)$ and absorbing boundary
$\Pi(\delta_c,S)=0$. This solution is
\begin{equation}
\Pi(\delta_c,\delta,S)=\frac{1}{\sqrt{2\pi S}}\left[e^{-\frac{\delta^2}{2S}}-e^{-\frac{(2\delta_c^2-\delta)^2}{2S}}\right],
\end{equation}
defined for $\delta<\delta_c$. Hence, the fraction of volume
occupied by halos, i.e. the fraction of trajectories which have
crossed the barrier ($\delta>\delta_c$) is given by
$F(S)=1-\int_{-\infty}^{\delta_c} \Pi(\delta_c,\delta,S)d\delta$. Evaluating
the first-crossing distribution $dF/dS$ and substituting in the
definition of $f(\sigma)$ we finally obtain 
\begin{equation}
f(\sigma)=\frac{\delta_c}{\sigma}\sqrt{\frac{2}{\pi}}\,e^{-\frac{\delta_c^2}{2\sigma^2}},
\end{equation}
that is the `Extended Press-Schechter' (EPS) mass function. 
In the next Sections we will discuss how 
a different choice of the halo collapse model 
and the smoothing function alter this standard result. 

\section{Halo Collapse and the diffusive drifting barrier}\label{barrmod}
The spherical collapse model \cite{GunnGott1972} provides a complete description of the non-linear
evolution of a spherically symmetric top-hat density perturbation embedded in a
Friedman-Lemaitre-Robertson-Walker background. 
A characteristic feature of this model is that the collapse 
does not depend on the initial size of the region, i.e. on the initial radius, but only on the 
amplitude of the initial top-hat overdensity. Since the enclosed mass depends
on the initial radius, then halos will be indistinctly associated to regions 
of the linear density field which lies above the same density
threshold.

The work of Doroshkevich \cite{Doroshkevich1970} has shown
that initial Gaussian density perturbations are highly non-spherical
and approximately triaxial. Hence, the collapse of 
a homogeneous ellipsoid (see e.g. \cite{EisensteinLoeb1995}) should provide a far better
description of the conditions of halo formation. In such a case the collapse depends on the initial
overdensity as well as the shear field and thus on the initial size of the ellipsoidal region, 
i.e. on the enclosed mass. Moreover, as also shown in
\cite{Doroshkevich1970}, because of the random nature of the density perturbation field,
the parameters characterizing the ellipsoid are random variables themselves with probability
distributions which depend on the statistics of the underlying density
fluctuation field. Consequently, the main feature of the non-spherical collapse
of halos is that the critical density threshold is a stochastic
variable itself. In the language of the Excursion Set this translates
into a `fuzzy' barrier performing a random walk whose properties 
depend on the specificity of the non-spherical collapse 
model considered (see e.g. \cite{Monaco1995,Audit1997,Lee1998}). 
For example, Sheth et al. \cite{Sheth2001} have shown that in the
ellipsoidal collapse model the barrier on average
evolves as $\langle B\rangle=\delta_c[1+\beta(S/\delta_c^2)^{\gamma}]$, with
$\beta=0.47$ and $\gamma=0.615$. This trend reflects the fact that in the 
small mass range (large $S$) there is a significant shear
that opposes the gravitational infall, hence a higher density
threshold is needed for the collapse to occur. Instead, for large masses
the shear field is less significant, halos are more isolated and the
collapse is approximately spherical. 

Moving barrier models have been
considered in a variety of case studies. Sheth \cite{Sheth1998} has
introduced a linear barrier model to study the halo distribution in
Eulerian space. In \cite{Shen2006} barriers with different functional dependence on $S$
where used to model the formation of halos, filaments and sheets,
while in \cite{furlanetto} a moving barrier has been introduce to
infer the size distribution of H II bubbles during reionization. These
studies have used Monte-Carlo simulations to infer the mass
distribution. Alternatively, Zhang \& Hui \cite{JunHui2006}
have introduced an integral-equation method which allows for a
recursive computation of the mass function for generic moving
barriers without requiring the use of Monte-Carlo simulations. Nevertheless, all
these studies have focused on rigid moving barriers rather than
stochastic ones. A realistic description of the non-spherical collapse
conditions of DM halos must necessarely account for this characteristic 
feature that, as we will see here, modifies the EPS result in a very distinctive way.
Furthermore, the analysis of numerical N-body simulations has indeed
confirmed the stochastic barrier hypothesis. For instance, the work by Robertson et
al. \cite{Robertson2009} has clearly shown that the linear density fluctuation associated with halos 
detected in the numerical simulations at a given redshift has not a
unique constant value, but is randomly distributed as function of the
halo mass, following an approximately log-normal distribution 
with a nearly linearly drifting average. 

Maggiore \& Riotto \cite{MaggioreRiotto2010b}
have implemented these features in the Excursion Set theory by assuming a Gaussian diffusive
barrier with $\langle B\rangle=\delta_c$ and $\langle(B-\langle
B\rangle)^2\rangle=S\,D_B$, where $D_B$ is a constant diffusion
coefficient which they set to $D_B\approx 0.3$ to reproduce the
results of \cite{Robertson2009}. 
\begin{figure}[bt]
\includegraphics[scale=0.4]{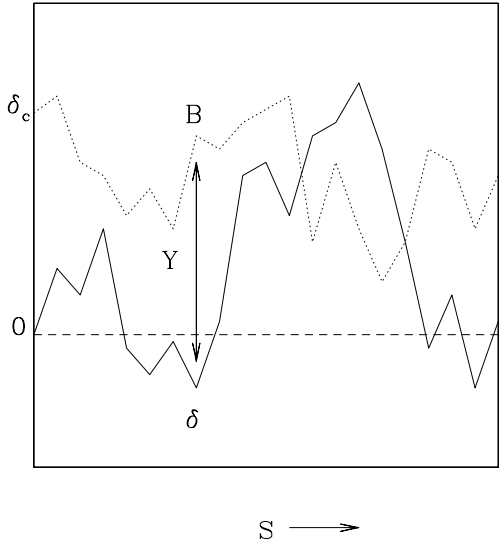}
\caption{Schematic representation of random walk trajectories for
  $\delta$ (solid line) and the barrier $B$ (dot line) as function of
  the variance. The variable $Y=B-\delta$ measure the distance between
  the two trajectories and therefore performs a random walk as well. 
  The trajectories starts at $\delta(0)=0$ and $B(0)=\delta_c$, i.e. $Y(0)=\delta_c$.}
\label{fig1}
\end{figure}
Here, we improve the modeling of the diffusive stochastic barrier
by assuming a linearly
drifting average, $\langle B\rangle\equiv\bar{B}(S)=\delta_c+\beta S$ (see
e.g. \cite{Sheth1998}), which approximates the average drift
predicted by the ellipsoidal collapse model in \cite{Sheth2001}. 
Under these assumptions the barrier obeys the following Langevin equation
\begin{equation}
\frac{\partial B}{\partial S}=\beta+\eta_{B}(S),\label{db}
\end{equation}
where the noise $\eta_B(S)$ is characterized by $\langle\eta_B(S)\rangle=0$ and 
$\langle\eta_B(S)\eta_B(S')\rangle=D_B\,\delta_D(S-S')$. Without loss
of generality we can assume that $\eta_B(S)$ and $\eta_\delta(S)$ are
uncorrelated and the stochastic evolution of the system is described by
Eqs.~(\ref{langevin_eta})-(\ref{db}) with absorbing boundary
at $\delta(S)=B(S)$. The system can be reduced to a one-dimensional
random walk by introducing the variable $Y=B-\delta$. In such a case
we have
\begin{equation}
\frac{\partial Y}{\partial S}=\beta+\eta(S),\label{dy}
\end{equation}
with white noise $\eta(S)=\eta_\delta(S)+\eta_B(S)$ such that $\langle \eta(S)\rangle=0$ and
$\langle \eta(S)\eta(S')\rangle=(1+D_B)\delta(S-S')$. 

In Fig.~\ref{fig1} we plot a schematic representation of the random
walks. The variable $Y$ measures the distance of the trajectory
of the density field to the barrier as function of the variance. 
The system starts at $S=0$ with $\delta(0)=0$ and $B(0)=\delta_c$, 
thus $Y(0)\equiv Y_0=\delta_c$. Random walks which have yet to form 
halos correspond to $Y>0$. The absorbing boundary is located at $Y=0$ 
and trajectories which have collapsed into halos correspond to $Y<0$. 
From Eq.~(\ref{dy}) we derive
the Fokker-Planck equation for the probability distribution $\Pi_{0}(Y_0,Y,S)$ of the random walks
with $Y>0$ (see e.g. \cite{VanKemp1992}):
\begin{equation}
\frac{\partial \Pi_{0}}{\partial S}=-\beta\frac{\partial \Pi_{0}}{\partial Y}+\frac{1+D_B}{2}\frac{\partial^2\Pi_{0}}{\partial Y^2},\label{FP}
\end{equation}
where the subscript `$0$' refers to the fact that the
random walks which we are considering are Markovian. 
We solve Eq.~(\ref{FP}) with initial condition $Y_0=\delta_c$ and impose the absorbing
boundary condition at $Y=0$, i.e. $\Pi_{0}(0,S)=0$. By rescaling the variable $Y\rightarrow \tilde{Y}=Y/\sqrt{1+D_B}$, 
a factorisable solution can be found in the form $\Pi_{0}(\tilde{Y},S)=U(\tilde{Y},S)\exp[c(\tilde{Y}-cS/2)]$, where 
$c=\beta/\sqrt{1+D_B}$ and $U(\tilde{Y},S)$ satisfies a Gaussian
diffusion equation which can be solved using the image method
\cite{Redner2001} or by Fourier transform. We find
\begin{equation}\Pi_{0}(Y_0,Y,S)=\frac{e^{\frac{\beta}{1+D_B}(Y-Y_0-\beta\frac{S}{2})}}{\sqrt{2\pi S(1+D_B)}}\left[e^{-\frac{(Y-Y_0)^2}{2S(1+D_B)}}-e^{-\frac{(Y+Y_0)^2}{2S(1+D_B)}}\right].\label{pimarkov}
\end{equation}

\begin{figure}[t]
\includegraphics[scale=0.42]{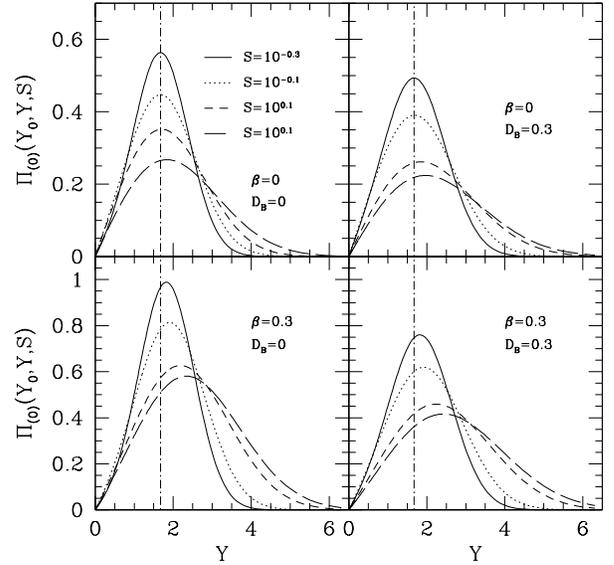}
\caption{Probability distribution of non-collapsing trajectories
  ($Y>0$) for $S=10^{-0.3}$ (solid line), $10^{-0.1}$ (dot line), 
$10^{0.1}$ (short dash line) and $10^{0.3}$ (long dash line). Top left
  panel: Extended Press-Schechter ($\beta=0$, $D_B=0$); top right
  panel: diffusive barrier ($\beta=0$, $D_B=0.3$); bottom left panel:
  linearly drifting barrier ($\beta=0.3$, $D_B=0$); bottom right
  panel: diffusive barrier with linearly drifting average ($\beta=0.3$, $D_B=0.3$).}
\label{fig2}
\end{figure}
 
Here, it is worth remarking that a general analytic solution to the Fokker-Planck
equation with biased diffusion and absorbing boundary condition 
does not exists for drift terms which are non-linear in $S$. As it
will be evident in the next Section, having an analytical expression for the probability distribution of the Markovian random
walk greatly simplify the computation of the non-Markovian
corrections induced by a realistic filtering of the linear density
field. For this very reason we have opted to assume a barrier model
with linearly drifting average, rather than the ellipsoidal collapse
prediction from \cite{Sheth2001}. 

In Fig.~\ref{fig2} we plot $\Pi_{0}(Y_0,Y,S)$ as function of $Y$
at $S=10^{-0.3}$ (solid line), $10^{-0.1}$ (dot
line), $10^{0.1}$ (short dash line) and $10^{0.3}$ (long dash line)
for different value of $\beta$ and $D_B$ such as to give us a
qualitative understanding of the barrier model parameter dependence. 
The standard EPS result corresponds to $\beta=0$ and
$D_B=0$ (top left panel). The case of a diffusive barrier with
$\beta=0$ and $D_B=0.3$ (as in \cite{MaggioreRiotto2010b}) 
is shown in the top right panel, while the
case of the linearly drifting average barrier with $\beta=0.3$ and
$D_B=0$ is shown in the bottom left\footnote{We find that for
  $\beta=0.3$ the linear drifting average barrier
  $\bar{B}(S)=\delta_c+\beta S$ approximates to better
than $10\%$ the prediction of the ellipsoidal collapse model
\cite{Sheth2001} over the range $-0.6<\log{(1/\sigma)}<0.4$.}.
Finally, the case of the diffusive barrier with linearly drifting average with
$\beta=0.3$ and $D_B=0.3$ is plotted in the bottom right panel. 

We may notice that overall amplitude of $\Pi_{0}(Y_0,Y,S)$ is a decreasing function
of $S$ with an increasing skewness toward larger values of $Y$. Since the total 
number of trajectories is conserved, this implies that the probability 
of trajectories that do not cross the barrier between $Y$ and $Y+dY$
decreases as function of $S$, while that of those which first-cross it
increases. This is consistent with that fact that in the bottom-up scenario
small mass halos are more likely to form than large ones. 
However, at a finer level the trend is barrier model dependent. For instance in
the case of the diffusive barrier (top right panel) we can see that
the amplitude of $\Pi_{0}(Y_0,Y,S)$ is smaller than
the standard EPS result (top left panel). Thus, indicating that the number
of crossing trajectories is higher.
In contrast, for the non-diffusive barrier with drifting average (bottom
left panel) we have that $\Pi_{0}(Y_0,Y,S)$ is larger than the EPS prediction. 
In addition, the peak of the probability distribution
rapidly shifts towards larger values of $Y$ as
function of $S$ as opposed to the EPS case. 
Finally, for the diffusing barrier 
with linearly drifting average (bottom right
panel) we may notice that the combined effect of diffusion and drift 
is to reduce the overall amplitude of $\Pi_{0}(Y_0,Y,S)$ more effectively
in the large mass range than in the low mass end.

The first-crossing distribution gives by definition
the probability $\mathcal{F}_0(S)$ of a random walk to cross 
the barrier between $S$ and $S+dS$, thus we have:
\begin{eqnarray}
\mathcal{F}_0(S)&=&-\frac{\partial}{\partial S}\int_0^{\infty}dY\,\Pi_{0}(Y_0,Y,S)\nonumber\\
&=&\beta\,\Pi_{0}(Y_0,Y,S)\bigg|_0^\infty-\frac{1+D_B}{2}\frac{\partial\Pi_{0}}{\partial
  Y}\bigg|_0^\infty\nonumber\\
&=&\frac{\delta_c}{S^{3/2}\sqrt{2\pi(1+D_B)}}e^{-\frac{(\delta_c+\beta S)^2}{2 S(1+D_B)}},
\end{eqnarray}
for $D_B=0$ this coincides with the non-diffusing linear drifting
barrier solution found in \cite{Zentner2007}. Then, the Markovian
mass function reads as:
\begin{equation}
f_0(\sigma)=\frac{\delta_c}{\sigma\sqrt{1+D_B}}\sqrt{\frac{2}{\pi}}\,e^{-\frac{(\delta_c+\beta\sigma^2)^2}{2\sigma^2(1+D_B)}}.\label{fsigmalin}
\end{equation}
The competing effect of diffusion and average drift, which we have inferred from
the qualitative analysis of $\Pi_{0}(Y_0,Y,S)$, can be seen directly
in the form of Eq.~(\ref{fsigmalin}). The non-vanishing diffusion
coefficient has the effect of reducing the amplitude of the mass
function cut-off, thus shifting it toward smaller values of $\sigma$. 
In contrast, the drift of the barrier
tends to increase the value of the threshold as function of $\sigma$. 

\begin{figure}[t]
\includegraphics[scale=0.42]{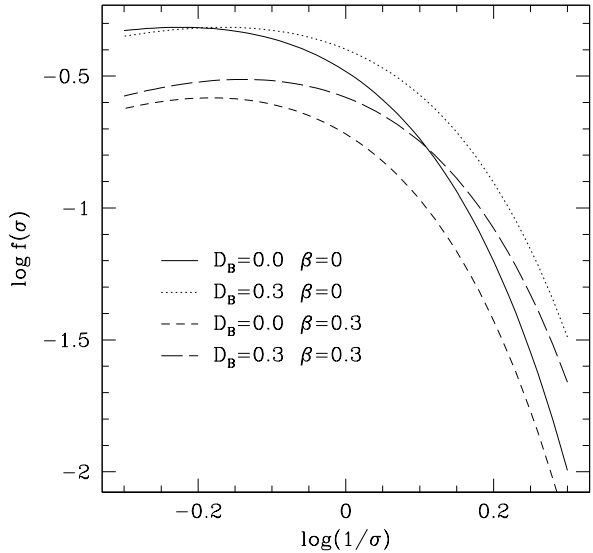}
\caption{Mass function for values of $\beta$ and $D_B$ as in Fig.~\ref{fig2}.}
\label{fig3}
\end{figure}

In Fig.~\ref{fig3} we plot $f_0(\sigma)$ for the same values of
$\beta$ and $D_B$ shown in Fig.~\ref{fig2}. The qualitative trend 
confirms that inferred from the analysis of
$\Pi_{0}(Y_0,Y,S)$. Firstly, we notice that with respect to the standard EPS case (solid
line), the diffusing barrier model (dot line) shows a cut-off at a
lower value of $\sigma$. This is because in the presence of diffusion
the condition of collapse has a scatter around the spherical collapse
threshold which favors the trajectory crossing. Consequently, the mass function
is larger than the EPS result over a larger range of masses.
In contrast, the linear drifting barrier
(short dash line) gives a mass function which is suppressed at all
masses. This is consistent with the fact that the barrier is on average higher than the spherical
collapse threshold, thus is more difficult for trajectories to cross
the barrier. In the case of a diffusive barrier with linearly drifting
average (long dash line) the combination of the two effects causes the mass
function to be tilted with respect to the EPS prediction. In the next
Section we will discuss the modifications to Eq.~(\ref{fsigmalin}) due
to a different choice of the smoothing function.
  
\section{Non-Markovian Corrections and Path-Integral Approach}\label{nonmark}

\subsection{Halo Mass Definition and Filter Function}
The filtering of the linear density fluctuation
field specifies the relation between the smoothing scale $R$ and the
mass $M$. The volume selected by the window function $W(x,R)$ is $V(R)=\int
d^3x\,W(x,R)$, hence the enclosed mass is given by $M(R)=\bar{\rho} V(R)$. 
However, this relation is uniquely specified
only for a sharp-x filter, $W(r,R)=\theta(r-R)$, for which $M(R)=4/3\pi\bar{\rho}
R^3$. For generic filters, the mass definition is ambiguous, since 
it is defined up to a normalization constant. More importantly,
in the case of the sharp-k filter, $W(k,R)=\theta(1/R-k)$, the mass 
remains undefined (see discussion in \cite{MaggioreRiotto2010a}).
On top of this, we should consider the fact that
the mass definition of N-body halos depends on the halo detection algorithm.
Thus, for a consistent model comparison with numerical
simulations, the filtering of the linear density field should be chosen 
consistently with the mass definition of the halo detection algorithm
used to measure the N-body mass function. 
As an example, the Spherical Overdensity (SOD) halo finder 
detects halos as spherical regions of radius $R_\Delta$ enclosing a 
density $\rho_\Delta=\Delta\bar{\rho}$, where $\Delta$
is the overdensity parameter usually fixed to $\Delta=200$ (which is
roughly equal to spherical collapse prediction of the virial overdensity at $z=0$ in LCDM models).
In such a case, the halo mass is $M_\Delta=4/3\pi\bar{\rho}\Delta
R_\Delta^3$, that is equivalent to definition of the sharp-x
filter. However, random walks generated by this smoothing function
are no longer Markovian. In fact, the Fourier transform of the sharp-x
filter is
\begin{equation}
\tilde{W}(k,R)=\frac{3}{(kR)^3}[\sin(kR)-(kR)\cos(kR)].\label{sharpx}
\end{equation}
and in such a case it is easy to see from Eq~(\ref{noisec}) that $\delta(S)$
is no longer subject to a simple white noise.

\begin{figure}[t]
\includegraphics[scale=0.42]{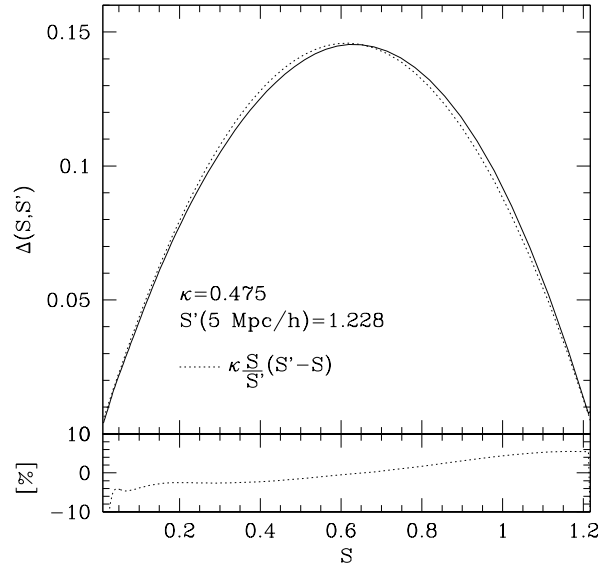}
\caption{Correlation function $\Delta(S,S')$ for a sharp-x filter (solid line) obtained by numerically integrating
Eq.~(\ref{corr}) for $S'(\textrm{R=5 Mpc/h})=1.228$ in the range $0<S<S'$, and the fitting function $\kappa\,S/S'(S'-S)$
with $\kappa=0.475$ (dash line).} 
\label{fig4}
\end{figure}

Maggiore \& Riotto \cite{MaggioreRiotto2010a} have shown that the
correlations induced by the filter function can be
treated as perturbations about the Markovian solution.
Let us consider a Gaussian random walk, the statistical properties 
are entirely specified by the $1$ and $2$-point connected
correlators. For a Gaussian field these are
$\langle\delta(R)\rangle_c=0$ and
\begin{equation}
\langle\delta(R)\delta(R')\rangle_c=\int_0^{\infty}\frac{k^2dk}{2\pi^2}P(k)\tilde{W}(k,R)\tilde{W}(k,R')\label{corr}
\end{equation}
respectively, with $P(k)=A\,k^{n_s}\,T^2(k)$ and $T(k)$ is the matter transfer
function. In the case of a sharp-k filter Eq.~(\ref{corr})
simply reduces to $\langle\delta(R[S])\delta(R[S'])\rangle_c=\textrm{min}(S,S')$.
However, Maggiore \& Riotto \cite{MaggioreRiotto2010a} have shown that for
the sharp-x filter the 2-point connect correlator can be written as
$\langle\delta(R[S])\delta(R[S'])\rangle_c=\textrm{min}(S,S')+\Delta(S,S')$,
where $\Delta(S,S')$ is a symmetric function which vanishes at
$S=S'$ and is well approximated by
$\Delta(S,S')\approx\kappa\,S/S'(S'-S)$ and 
$\kappa$ is a fitting coefficient. Following their computation, we estimate $\Delta(S,S')$
in a $\Lambda$CDM cosmology with model parameters set to the WMAP-5yr
best fit values: 
$\Omega_M=1-\Omega_\Lambda=0.28$, $h=0.7$, $\sigma_8=0.8$ and $n_s=0.96$. We compute the transfer function using the
CMBFAST code \cite{SeljakZaldarriaga1996}. In Fig.~(\ref{fig4}) we plot $\Delta(S,S')$ for $S'(\textrm{R=5 Mpc/h})=1.228$ 
(solid line) against $\kappa\,S/S'(S'-S)$ for the best fit value
$\kappa=0.475$. We can see that difference 
between the numerically evaluated correlation function and the best
fitting formula is well within a few per-cent level. Note that $\kappa=0.475$
slightly differs from the value found in \cite{MaggioreRiotto2010a}. The discrepancy is due to the difference  
between the numerically computed CMBFAST transfer function and the approximate 
fitting formula by Sugiyama \cite{Sugiyama1995} which has been used in
\cite{MaggioreRiotto2010a}. The coefficient
$\kappa$ depends on the assumed cosmological model and does not evolve in
redshift. As pointed out in \cite{MaggioreRiotto2010a}, a weakly
linear dependence of $\kappa$ on the smoothing scale $R$ can improve
the fit to the numerically computed correlation function
$\Delta(S,S')$. Nevertheless, such dependece can be neglected to first approximation.

In terms of the variable $Y$ the non-vanishing connected correlators
read as $\langle Y(S)\rangle_c\equiv\bar{B}(S)=\delta_c+\beta S$ and $\langle
Y(S)Y(S')\rangle_c=(1+D_B)\textrm{min}(S,S')+\Delta(S,S')$ respectively. 
Having assumed the barrier to perform Gaussian random walks implies that the non-Markovian
part of the 2-point correlator identically vanishes, i.e. $\Delta_B(S,S')=0$.
One may wonder whether a term like $\Delta_B(S,S')$ due to the filter function
should also be included. However, there is no reason as to why the barrier 
has to have the same filtering of the linear density field, since the two smoothing procedures
have very different physical meanings. The latter is related to the halo mass definition, 
while the former specifies the correlation between the condition of halo collapse at different
scales. The simplest approximation is to assume that the 
collapse at a scale $S$ is independent of that at $S'$, which is 
equivalent to having $\Delta_B(S,S')=0$, i.e. Gaussian random walks. 
This suggests that a variety of non-linear gravitational effects 
which induce scale-dependent correlations of the halo collapse condition 
can be implemented in such a formalism through the barrier $p$-point 
connected correlators with $p\ge 2$ or using a non-trival smoothing procedure
of the barrier random walks.

\subsection{Path-Integral Method}\label{pathint}
Hereafter, we will follow 
the derivation of \cite{MaggioreRiotto2010a},
and extend the computation of the non-Markovian corrections the case of the diffusive barrier model
with linearly drifting average. 

Let us consider the random walk
of the variable $Y$ over the time interval $[0,S]$ discretized in steps
$\Delta S=\epsilon$, such that $S_k=k\epsilon$ with $k=1,..,n$.
The probability distribution of trajectories that start at $Y_0$,
end in $Y_n$ at time $S_n$ and which never cross the
barrier is given by
\begin{equation}
\Pi_\epsilon(Y_0,Y_n,S_n)=\int_0^\infty dY_1\,...\int_0^\infty dY_{n-1} W(Y_0,..,Y_n,S_n),\label{pepsilon}
\end{equation}
where
\begin{equation}
W(Y_0,..,Y_n,S_n)\equiv\langle\delta_D(Y(S_1)-Y_1)...\delta_D(Y(S_n)-Y_n)\rangle,
\end{equation}
is the probability density distribution. Using the Fourier transform
of the Dirac-function we have
\begin{equation}
W(Y_0,..,Y_n,S_n)=\int\mathcal{D}\lambda\, 
  e^{i\sum_{i=1}^n\lambda_i Y_i}\langle e^{-i\sum_{i=1}^n\lambda_i Y(S_i)}\rangle,
\end{equation}
with $\int\mathcal{D}\lambda=\int_{-\infty}^\infty \frac{d\lambda_1}{2\pi}...\frac{d\lambda_{n}}{2\pi}$. 
The exponential factor within the brackets can be written in
terms of the connected correlators $\langle
Y(S)\rangle_c$ and $\langle Y(S)Y(S')\rangle_c$
\begin{equation}
\langle e^{-i\sum_{i=1}^n\lambda_i
  Y(S_i)}\rangle=e^{-i\sum_{i=1}^n\lambda_i\bar{B}_i-\frac{1}{2}\sum_{ij}\lambda_i\lambda_j(A_{ij}+\Delta_{ij})}\label{wdensity}
\end{equation}
where we have indicated with $\bar{B}_i=\bar{B}(S_i)$,
$A_{ij}=(1+D_B)\textrm{min}(i,j)\epsilon$ and $\Delta_{ij}=\kappa
S_i/S_j(S_j-S_i)$. Substituting Eq.~(\ref{wdensity}) in Eq.~(\ref{pepsilon}) and 
given the fact that $\Delta_{ij}<1$ we can expandi in $\kappa$, to first order we have 
\begin{widetext}
\begin{equation}
\Pi_\epsilon(Y_0,Y_n,S_n)=\int_0^\infty dY_1\,...\int_0^\infty dY_n \int \mathcal{D}\lambda \left(1-\frac{1}{2}\Sigma_{ij}\lambda_i\lambda_j\Delta_{ij}\right)e^{-i\Sigma_k\lambda_k [\bar{B}_k-Y_k]}e^{-\frac{\epsilon}{2}\Sigma_{ij}\lambda_i\lambda_j\tilde{A}_{ij}},\label{pitotepsilon}
\end{equation}
\end{widetext}
where $\tilde{A}_{ij}=(1+D_B)\textrm{min}(i,j)$. Thus
\begin{equation}
\Pi_\epsilon(Y_0,Y_n,S_n)=\Pi_{0}^\epsilon(Y_0,Y_n,S_n)+\Pi_{1}^\epsilon(Y_0,Y_n,S_n), 
\end{equation}
where $\Pi_{0}^\epsilon(Y_0,Y_n,S_n)$ is the Markovian zero order term and 
$\Pi_{1}^\epsilon(Y_0,Y_n,S_n)$ is the first order correction in
$\kappa$. The Markovian term obeys the Chapman-Kolmogorov equation
(see Eq.~(\ref{chapman}) in Appendix \ref{app0} for a detailed
derivation) which is used to explicitely compute the non-Markovian term.

A crucial point concerns the $\epsilon$-dependence 
of $\Pi_{0}^\epsilon(Y_0,Y_n,S_n)$ in the proximity of the barrier.
As extensively discussed in \cite{MaggioreRiotto2010a}, the Markovian solution 
$\Pi_{0}^\epsilon(Y_0,Y_n,S_n)$ is $\mathcal{O}({\epsilon})$ for
$Y_n>0$ and $\mathcal{O}({\epsilon}^{1/2})$ at $Y_n=0$. Hence the
probability distribution undergoes a transition between two different
regimes inside a boundary layer of finite size. In order to 
evaluate the form of the probability distribution inside this region 
it is convenient to introduce the `stretch' variable
$\eta=Y/\sqrt{2\epsilon(1+D_B)}$ and write 
\begin{equation}
\Pi_{0}^\epsilon(Y_0,Y_n,S_n)=C_\epsilon(Y_0,Y_n,S_n)u(\eta),\label{nearb}
\end{equation}
where $C_\epsilon(Y_0,Y_n,S_n)$ is a smooth solution and $u(\eta)$ is
a function containing the fast variation in $\epsilon$ inside the
transition region. In the continuous
limit, $\lim_{\eta\rightarrow \infty}u(\eta)=1$,
while $C_\epsilon$ tends to
Eq.~(\ref{pimarkov}), thus we recover the standard Markovian
solution. Substituting $Y\equiv Y_n=\eta\sqrt{2\epsilon(1+DB)}$ in Eq.~(\ref{pimarkov}) and 
expanding to lowest order in $\epsilon$ we obtain
\begin{equation}
C_\epsilon(Y_0,Y_n,S_n)=\sqrt{\epsilon}\frac{2 \eta Y_0}{\sqrt{\pi}S_n^{3/2}(1+D_B)}e^{-\frac{(Y_0+\beta S_n)^2}{2 S_n(1+D_B)}}.\label{cesp}
\end{equation}
Hence, substituting Eq.~(\ref{cesp}) in Eq.~(\ref{nearb}) and taking
the limit $\eta\rightarrow 0$ we obtain
\begin{equation}
\Pi_{0}^\epsilon(Y_0,0,S_n)=\sqrt{\epsilon}\frac{\gamma Y_0}{S_n^{3/2}(1+D_B)}e^{-\frac{(Y_0+\beta S_n)^2}{2 S_n(1+D_B)}},\label{atb}
\end{equation}
where 
\begin{equation}
\gamma\equiv\frac{2}{\sqrt{\pi}}\lim_{\eta\rightarrow 0}\eta\,u(\eta)=\frac{1}{\sqrt{\pi}},\label{gamma}
\end{equation}
we present the exact derivation of this result in Appendix \ref{app1}. 

Similarly we can infer the probability $\Pi_{0}^\epsilon(0,Y_n,S_n)$
of trajectories starting at the barrier $Y_0=0$ and ending at $Y_n>0$
by introducing the stretch variable $\eta=Y_0/\sqrt{2\epsilon(1+D_B)}$.
Again, substituting $Y_0=\eta\sqrt{2\epsilon(1+D_B)}$ in Eq.~(\ref{pimarkov}),
expanding to lowest order in $\epsilon$ and computing
Eq.~(\ref{nearb}) in the limit $\eta\rightarrow 0$ we obtain
\begin{equation}
\Pi_{0}^\epsilon(0,Y_n,S_n)=\sqrt{\epsilon}\frac{\gamma Y_n}{S_n^{3/2}(1+D_B)}e^{-\frac{(Y_n-\beta S_n)^2}{2 S_n(1+D_B)}}.\label{fromb}
\end{equation}

Finally, the probability of trajectories which start at the barrier
and end at the barrier can be obtained using the dimensional
arguments discussed in \cite{MaggioreRiotto2010a}. To lowest order in
$\epsilon$, we find
\begin{equation}
\Pi_{0}^\epsilon(0,0,S_n)=\epsilon\frac{1}{S_n^{3/2}\sqrt{2\pi(1+D_B)}}.\label{fromtob}
\end{equation}

\subsection{Non-Markovian Corrections to the Halo Mass Function}
We have now all the ingredients to compute the non-Markovian correction to first order in $\kappa$.
Using the fact that
$\lambda_i e^{-i\lambda_i [\bar{B}_i-Y_i]}=-i\partial/\partial{Y_i}
e^{-i\Sigma_{ij}\lambda_k[\bar{B}_k-Y_k]}$, the second term in
Eq.~(\ref{pitotepsilon}) reads as
\begin{eqnarray}
\Pi_{1}^\epsilon(Y_0,Y_n,S_n)&=&\frac{1}{2}\sum_{ij}\int_{0}^\infty dY_1...\int_0^\infty
dY_{n-1}\,\Delta_{ij}\times\nonumber\\
&\times&\partial_i\partial_j W_0(Y_0,..,Y_n,S_n),\label{nmcepsilon}
\end{eqnarray}
where $W_0(Y_0,..,Y_n,S_n)$ is the Markovian probability density of
the random walks (see Appendix \ref{app0}). We split the sum in
Eq.~(\ref{nmcepsilon}) as
\begin{equation}
\frac{1}{2}\sum_{ij}\Delta_{ij}\partial_i\partial_j=\frac{1}{2}\Delta_{nn}\partial_n^2+\sum_{i=1}^{n-1}\Delta_{in}\partial_i\partial_n+\sum_{i<j}\Delta_{ij}\partial_i\partial_j,\label{sum}
\end{equation}
notice that the first term on the right-hand side
of Eq.~(\ref{sum}) vanishes since $\Delta_{nn}=0$, furthermore 
$\sum_{i<j}=\sum_{i=1}^{n-2}\sum_{j=i+1}^{n-1}$. Then, 
integrating Eq.~(\ref{nmcepsilon}) by parts and using
Eq.~(\ref{chapman}) we obtain
\begin{equation}
\Pi_{1}^\epsilon(Y_0,Y_n,S_n)=\Pi^{\textrm{m}}_{\epsilon,1}(Y_0,Y_n,S_n)+\Pi^{\textrm{m-m}}_{\epsilon,1}(Y_0,Y_n,S_n),
\end{equation}
where
\begin{equation}
\begin{split}
\Pi^{\textrm{m}}_{\epsilon,1}(Y_0,Y_n,S_n)&=-\sum_{i=1}^{n-1}\Delta_{in}\partial_n\biggl[\Pi_{0}^\epsilon(Y_0,0,S_i)\times\\
&\times\,\Pi_{0}^\epsilon(0,Y_n,S_n-S_i)\biggr],\label{mem}
\end{split}
\end{equation}
is a `memory'-like term, since it depends on a single sum over the
past time steps, and 
\begin{equation}
\begin{split}
\Pi^{\textrm{m-m}}_{\epsilon,1}(Y_0,Y_n,S_n)=\sum_{i=1}^{n-2}\sum_{j=i+1}^{n-1}\Delta_{ij}\biggl[\Pi_{0}^\epsilon(Y_0,0,S_i)\times\\
\times\,\Pi_{0}^\epsilon(0,0,S_j-S_i)\,\Pi_{0}^\epsilon(0,Y_n,S_n-S_j)\biggr],\label{memofmem}
\end{split}
\end{equation}
which represents a `memory-of-memory' term. For a detailed derivation
of these equations see Appendix~\ref{app2}. The probability
distributions in Eq.~(\ref{mem}) and (\ref{memofmem}) are given by
Eq.~(\ref{atb}), (\ref{fromb}) and (\ref{fromtob}) respectively. 
In order to compute the sum over the time steps we take the continous
limit such that
\begin{equation}
\begin{array}{lr}
\sum_{i=1}^{n-1}\rightarrow \frac{1}{\epsilon}\int_0^{S}dS_i\,\,\mbox{and} &
\sum_{i<j}\rightarrow \frac{1}{\epsilon^2}\int_0^{S}dS_i\int_{S_i}^{S}dS_j.\nonumber
\end{array}
\end{equation}

We find that the `memory' term not to contribute to the mass
function. As a result of the integration over $dS_i$ we have 
\begin{equation}
\begin{split}
\Pi^{\textrm{m}}_1&(Y_0,Y,S)=-\frac{\kappa
  Y_0}{(1+D_B)^2}\times\\
&\times\frac{\partial}{\partial
  Y}\biggl\{Y\,e^{\frac{\beta}{1+D_B}\left(Y-Y_0-\beta\frac{S}{2}\right)}\,\textrm{Erfc}\biggl[\frac{Y_0+Y}{\sqrt{2 S (1+D_B)}}\biggr]\biggr\},\label{pimmint}
\end{split}
\end{equation}
and since the first-crossing distribution is given by $\mathcal{F}_1^{m}=-\frac{\partial}{\partial
  S}\int_0^\infty dY \,\Pi^{\textrm{m}}_1(Y_0,Y,S)$, the subsequent integration
of Eq.~(\ref{pimmint}) over $dY$ vanishes. This is
consistent with the result of Maggiore \& Riotto
\cite{MaggioreRiotto2010a}.

\begin{figure}[t]
\includegraphics[scale=0.42]{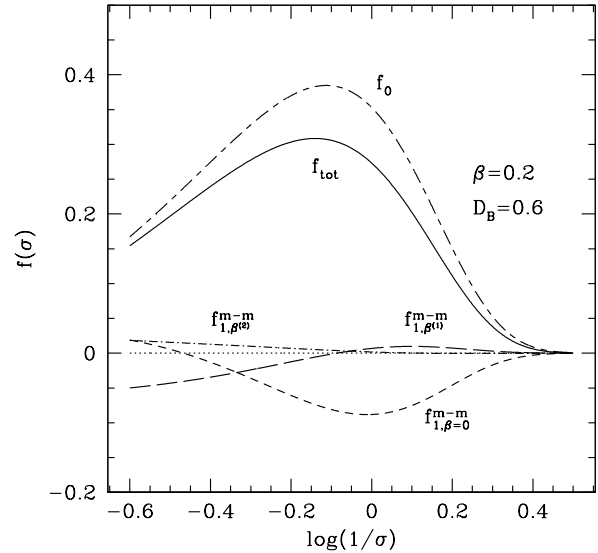}
\caption{Contributions to the halo mass function $f_\textrm{tot}$
 (solid line) for $\beta=0.2$ and $D_B=0.6$. The different curves 
correspond to the Markovian mass function $f_0$ (dot line) and
the non-Markovian corrections $f^{\textrm{m-m}}_{1,\beta=0}$
(short dash line), $f^\textrm{m-m}_{1,\beta^{(1)}}$ (long dash line),
$f^{\textrm{m-m}}_{1,\beta^{(2)}}$ (dot-short dash line),
$f^{\textrm{m-m}}_{1,\beta^{(3)}}$ (dot-long dash line).}
\label{fig5}
\end{figure}

The `memory-of-memory' term cannot be computed
analytically. Nevertheless, from the ellipsoildal collapse
model we have $\beta<1$. Thus, we can expand the integrands in powers of $\beta$ and compute
the contribution to the mass function up to leading order.
We find
\begin{equation}\label{f1b0}
f_{1,\beta=0}^{m-m}(\sigma)=-\tilde{\kappa}\frac{\delta_c}{\sigma}\sqrt{\frac{2a}{\pi}}\left[e^{-\frac{a\delta_c^2}{2\sigma^2}}-\frac{1}{2}\Gamma\left(0,\frac{a\delta_c^2}{2\sigma^2}\right)\right],
\end{equation}
where $a=1/(1+D_B)$, $\tilde{\kappa}=\kappa\,a$ and $\Gamma(0,z)$ is the incomplete Gamma function. Not surprisingly this expression coincides with the 
$\kappa$-correction for the diffusive barrier obtained in
\cite{MaggioreRiotto2010b}. The first order in $\beta$ reads as
\begin{equation}\label{f1b1}
f_{1,\beta^{(1)}}^{m-m}(\sigma)=-\beta\,a\,\delta_c\left[f_{1,\beta=0}^{m-m}(\sigma)+\tilde{\kappa}\,\textrm{Erfc}\left(\frac{\delta_c}{\sigma}\sqrt{\frac{a}{2}}\right)\right],
\end{equation}
while the second order is given by
\begin{equation}
\begin{split}
&f_{1,\beta^{(2)}}^{m-m}(\sigma)=\beta^2\,a\,\delta_c\,\tilde{\kappa}\biggl\{a\,\delta_c\,\textrm{Erfc}\left(\frac{\delta_c}{\sigma}\sqrt{\frac{a}{2}}\right)+\\
&+\sigma\sqrt{\frac{a}{2\pi}}\biggl[e^{-\frac{a\delta_c^2}{2\sigma^2}}\left(\frac{1}{2}-\frac{a\delta_c^2}{\sigma^2}\right)+\frac{3}{4}\frac{a\delta_c^2}{\sigma^2}\Gamma\left(0,\frac{a\delta_c^2}{2\sigma^2}\right)\biggr]\biggr\}.
\end{split}
\end{equation}
Higher order corrections can be computed semi-analytically, since
these contain integrals that cannot be written in terms of basic
functions. Simple numerical integration routines are sufficient to
evaluate these integrals. Nevertheless, we find that for
$\beta/(1+D_B)<1$ terms $\mathcal{O}(>\beta^2)$ are negligible. For example in Fig.~\ref{fig5} we plot the 
Markovian mass function $f_0$ and the $\kappa$-corrections up to $\mathcal{O}(\beta^3)$
and their total contribution for $\beta=0.2$ and $D_B=0.6$. We can see
that the term $\mathcal{O}(\beta^3)$ is negligible. The largest non-Markovian corrections
are given by $f_{1,\beta=0}^{m-m}(\sigma)$ in the
intermediate and high mass range, and $f_{1,\beta^{(1)}}^{m-m}(\sigma)$ 
at low masses. The overall effect of these non-Markovian terms is to reduce
the overall amplitude of the Markovian solution. 

Notice that the non-Markovian correction $f_{1,\beta=0}^{m-m}(\sigma)$ diverges in the very low mass
limit ($\sigma\rightarrow \infty$) due to the behavior of the
incomplete Gamma function in Eq.~(\ref{f1b0}). As shown in Fig.~\ref{fig5}, this term
(short dash line) decreases to a negative
minimum value at $\log(1/\sigma)\sim 0$ and then increases in the low mass range,
eventually diverging at very small masses, $\log(1/\sigma)\ll-0.6$
(corresponding to $M\ll 10^{10}\textrm{h}^{-1}\textrm{M}_\odot$). 
This implies that the mass function in the case of a diffusive barrier
with constant average is ill behaved in the very small mass limit. Hence, 
in order to extend its validity one should consider the contribution of 
higher-order corrections in $\kappa$. However, as we have previously
discussed, the low mass end of the mass function is sensitive to the
non-spherical collapse of halos. In such a case the non-Markovian
correction due to the average drift of the barrier to leading order in
$\beta$, $f_{1,\beta^{(1)}}^{m-m}(\sigma)$, cures the divergent behavior
of $f_{1,\beta=0}^{m-m}(\sigma)$. This can be inferred directly
from Eq.~(\ref{f1b1}), where the dependence on the
incomplete Gamma function is of opposite sign to that in
Eq.~(\ref{f1b0}). As shown in Fig.~\ref{fig5}, $f_{1,\beta^{(1)}}^{m-m}(\sigma)$ (long dash line) decreases towards negative
values as $f_{1,\beta=0}^{m-m}(\sigma)$ increases towards positve
values for small values of $\log(1/\sigma)$. 
The dependence on the Gamma function also appears in
$f_{1,\beta^{(2)}}^{m-m}(\sigma)$, however this correction is
subdominant with respect to $\mathcal{O}(\beta)$. Eventually, one can
find a small mass range where higher-order corrections in $\kappa$ and $\beta$
cannot be neglected. However, our analysis shows that in the mass range probed by current
simulations, $-0.6<\log(1/\sigma)<0.4$ (corresponding
$10^{10}<M[\textrm{h}^{-1}\textrm{M}_\odot]<10^{15}$)
the perturbative expansion in $\kappa$ and $\beta$ is well behaved
provided $\beta/(1+D_B)<1$.

\section{Excursion Set Mass Function and N-body Simulations}\label{numan}
The barrier model which we have considered here aims to capture the main features
of the ellipsoidal collapse of dark matter halos. It explicitly
depends on $\beta$ and $D_B$ which parametrize the
properties of the collapse threshold. In principle, these
parameters can be determined for a given ellipsoidal collapse
model. This is because the distribution of collapse density values
is directly related to the probability distribution of the eigenvalues
of the deformation tensor (see e.g. \cite{BondMyers1996,Sheth2001}). 
Alternatively, one can infer such a distribution by 
numerically solving the ellipsoidal collapse equations for randomly
generated initial conditions (see e.g. \cite{ChiuehLee,ShethTormen2002,Desjacques2008}).
Then, the values of $\beta$ and $D_B$ can be inferred by
best fitting the average and the variance of the inferred ellipsoidal
collapse density distribution. 

Most of the works in the literature have primarely focused on
determining the average of the ellipsoidal collapse threshold
\cite{Sheth2001,Shen2006}, while no attention has been
paid to the variance. In \cite{MaggioreRiotto2010b} the authors
have provided a rough estimate of the variance from the ellipsoidal collapse
barrier numerically determined in \cite{Sandvik2007}, though 
they did not use such estimate when evaluating the mass function. 
Furthermore, it is very plausible that the distribution collapse densities varies with cosmology and redshift. 
Hence, an accurate ellipsoidal collapse model prediction of $\beta$
and $D_B$ requires a dedicated study which is beyond the scope 
of this paper. 

Here, we test whether the path-integral inferred mass function can provide 
a reasonable description of the numerical simulation data.
In order to perform such a test we use the measurements of the halo mass function
from Tinker et al. \cite{Tinker2008} obtained using SOD with
$\Delta=200$ on a set of WMAP-1yr and WMAP-3yr cosmological N-body
simulations. First, we consider the mass function measurements at
$z=0$. For the LCDM models best fitting WMAP-1yr and 3yr data the
spherical collapse model prediction is $\delta_c=1.673$. Using such a
value we run a likelihood Markov Chain Monte Carlo analysis of the excursion set
mass function $f_{\textrm{tot}}(\sigma)$ including non-Markovian corrections up to
$\mathcal{O}(\beta^3)$ against the data to infer the best fit values of $\beta$
and $D_B$. The prior parameter space is $\log{\beta}=[-4,0]$ and $\log{D_B}=[-3,0]$. 

\begin{figure}[t]
\includegraphics[scale=0.42]{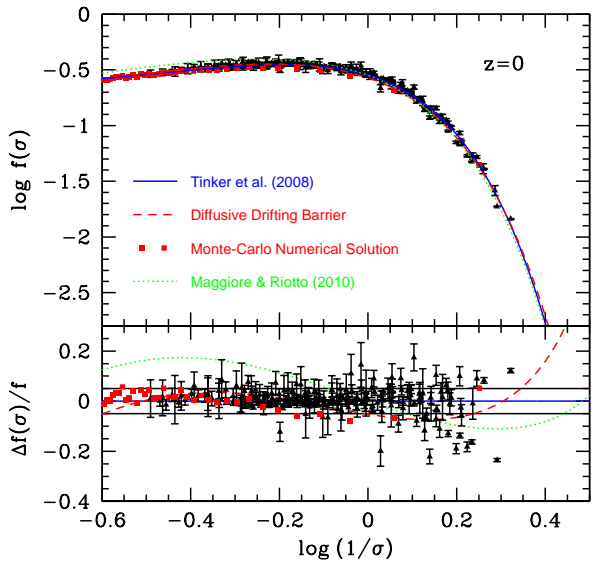}
\caption{(Upper panel) Halo mass function at $z=0$ for the Tinker et
  al. fitting formula with $\Delta=200$ (solid blue line), 
diffusing drifting barrier with ${\beta}=0.057$ and ${D}_b=0.294$ (red dash line) and the corresponding numerical solution from Monte Carlo generated random walks with sharp-x filter (red squares), Maggiore \& Riotto 
\cite{MaggioreRiotto2010b} with $D_B=0.235$ (green dot line). Data points are from \cite{Tinker2008}. 
(Lower panel) Relative difference with respect to the Tinker et
al. \cite{Tinker2008} fitting formula. The thin black solid lines indicates $5\%$ deviations.}
\label{fig6}
\end{figure}

We find ${\beta}=0.057$ and ${D}_B=0.294$ respectively. The numerical
simulation data strongly constrain these parameters 
with $1\sigma$ errors $\sigma_{\beta}=0.001$ and $\sigma_{D_B}=0.001$ respectively. 
We have also verified that these results do not change
if corrections $\mathcal{O}(>\beta^3)$ are included in
$f_{\textrm{tot}}(\sigma)$. In Fig.~\ref{fig6} (upper panel)
we plot the best fitting mass function (red dash line) against the simulation data 
together with the four-parameter fitting formula by Tinker et al. \cite{Tinker2008} for $\Delta=200$ (solid blue line). 
For comparison we also plot the diffusive barrier by Maggiore \&
Riotto \cite{MaggioreRiotto2010b} best fitting the data with
$D_B=0.235$ (green dot line). 
In Fig.~\ref{fig6} (lower panel) we plot the relative differences with respect to the Tinker et al. formula.
We may notice the remarkable agreement of the diffusive drifting barrier with the data. Deviations with 
respect to Tinker et al. (2008) are within the $\approx 5\%$ level for
$\log(1/\sigma)<0$ and within $7\%$ over the range $0<\log(1/\sigma)<0.3$. This
is quite impressive given the fact that our model depends only
on two physical parameters.
As expected the improvement with respect to the diffusive barrier
\cite{MaggioreRiotto2010b} is due to the drifting average which systematically
suppresses the formation of small mass halos with respect to the
massive ones.
In Fig.~\ref{fig6} we also plot the mass function for the diffusive
barrier model with ${\beta}=0.057$ and ${D}_b=0.294$ inferred from
Monte-Carlo generated random walks with sharp-x filtering (red
squares). As it can be seen this numerical solution well reproduce our
mass function formula with the same level of deviations from 
the Tinker et al. data\footnote{Recently, in \cite{Paranjape2011} the authors claim that the path-integral approach 
does not reproduce the Monte-Carlo inferred mass function to no better than $20\%$ and thus
conclude that such a formalism is inappropriate to fit halo 
abundances $dn(M)/dM$. Our numerical evaluation 
for the diffusive barrier model with linearly drifting
average shown in Fig.~\ref{fig6} clearly demonstrates this not to be the case.}. 

It is worth noticing that the best fit value of $\beta$ is about a
factor $5$ smaller than the ellipsoidal collapse model expectation.
This could be an artifact of the SOD mass function measurements, a consequence of 
our modeling of the barrier diffusion as a Gaussian random walk rather
than log-normal, or having limited the non-Markovian
corrections to first order in $\kappa$. In a future study we will 
perform a more detailed analysis to discriminate between these 
different effects. This will allow us to obtain an unbiased physical 
interpretation of the barrier model parameter values. 

\begin{figure}[t]
\includegraphics[scale=0.42]{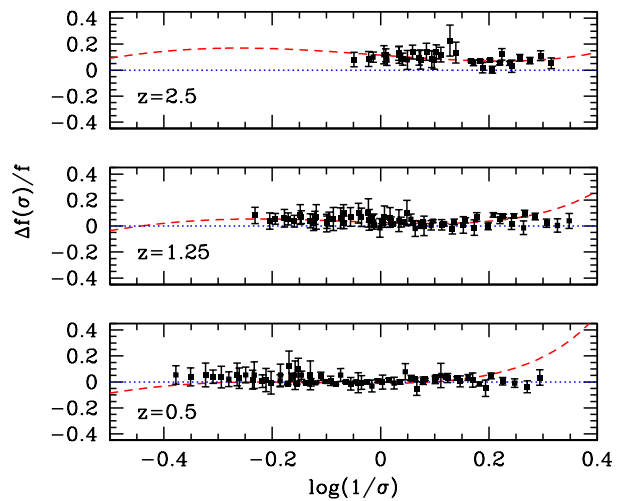}
\caption{Redshift evolution of the mass function residuals with
  respect to the redshift dependent Tinker et al. \cite{Tinker2008} fitting function (blue solid
  line) for diffusive barrier with linearly drifting average (red
  dash line).}
\label{fig8}
\end{figure}
\begin{figure}[t]
\includegraphics[scale=0.42]{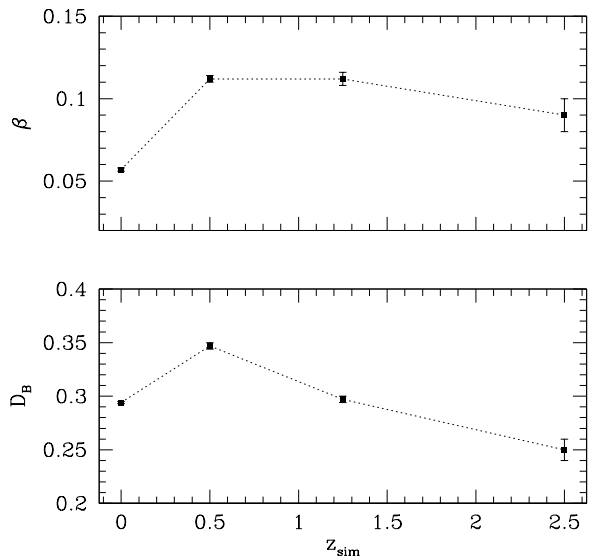}
\caption{Best fitting values of $\beta$ (upper panel) and $D_B$ (lover
  panel) as function of redshift. }
\label{fig7}
\end{figure}

The mass function measurements from Tinker et al. \cite{Tinker2008}
extend to $z=0.5,1.25$ and $2.5$. The mass function
can also reproduce these measurements for a given combination of
values of $\beta$ and $D_B$. At these redshifts we have 
$\delta_c=1.680,1.685$ and $1.686$ respectively \cite{Courtin2010}. 

In Fig.~\ref{fig8} we plot the residual of the mass
function $f_{\textrm{tot}}(\sigma)$ for the barrier model parameters
best fitting the data at $z=0.5,1.25$ and $2.5$ with respect to redshift dependent
fitting formula by Tinker et al. \cite{Tinker2008}. We can see that,
apart a systematic deviation at high masses ($\log(1/\sigma>0.2)$) of
order of $\lesssim 10\%$, $f_{\textrm{tot}}(\sigma)$
is consistent with data to better than $5\%$. In particular at $z=2.5$ the functional form
of the halo mass function seems to better reproduce the numerical
measurements compared to the Tinker et al. \cite{Tinker2008} fitting formula for which the authors
have found residuals $>5\%$ at $z=2.5$. In Fig.~\ref{fig7} we plot the best fit
values of $\beta$ and $D_B$ as function of $z$. The value of $\beta$
tends to saturate at $z>0.5$, which would imply that the non-spherical
collapse threshold on average has a similar mass (scale) dependence at
higher redshifts than at present. On the other hand, $D_B$ peaks at $z=0.5$ and then
decreases. Since, the diffusion coefficient primarily affect the
high-mass end of the mass function, this would suggest that the
collapse threshold of massive halos at higher redshifts is closer
to that predicted by the spherical collapse.
However, because of the limited z-sampling as well as a
systematic bias of the mass function data towards high masses the
physical interpretation of these trends should be taken carefully.
We leave a detailed analysis of these dependencies to a future numerical 
study.

\section{Linear Halo Bias}\label{secbias}
Halos are biased tracers of the dark matter density perturbations from
which they form. Operationally the halo bias is defined as the ratio of
the 2-point halo spatial correlation function to that of the underlying dark matter density
fluctuation field. In the Excursion Set formalism this can be
estimated using the peak-background split technique 
(see e.g. \cite{Bardeenetal1986,Cole&Kaiser1989,Mo&White1996,ShethTormen1999})
The basic idea is to evaluate the conditional first-crossing distribution and infer
the relative abundance of halos of a given mass (i.e. $S$) as function of the large scale density
fluctuation $\delta_0$, $\mathcal{F}(S|\delta_0,S'=0)$.
Then, it can be shown that to first order in $\delta_0$ the halo bias is given by
\begin{equation}\label{bias}
b_\textrm{h}(S)=1+\frac{1}{\mathcal{F}(S|0,0)}\frac{\partial\mathcal{F}(S|\delta_0,0)}{\partial \delta_0}\bigg|_{\delta_0=0},
\end{equation}
where $\mathcal{F}(S|0,0)$ coincides with the unconditional first-crossing distirbution with $\delta_0=0$. 

In the framework of the Excursion Set theory the halo bias for the
sharp-x filter with the non-Markovian corrections to first order in $\kappa$
has been derived in \cite{Ma2011}. We extend their calculation 
to the diffusive barrier with linearly drifting average (see also
\cite{Zhang2008} for the case with sharp-k filter and 
\cite{DeSimone2011} for a computation of the conditional mass function
in the case of generic moving barrier models and sharp-x filter). 
The calculations are quite cumbersome and since the basic results by Ma et
al. \cite{Ma2011} applies also to our case, we will report only the
relevant passages.

In order to evaluate $\mathcal{F}(S|\delta_0,0)$ let us first
compute the conditional mass function with conditioning on a generic scale $S'<S$ where
$\delta(S')\ll B(S')$. We find convenient to work with the variable
$Y=B-\delta$, even though the conditional first-crossing distribution 
$\mathcal{F}(S|\delta',S')$ differs from
$\mathcal{F}(S|Y',S')$. In fact, the latter imposes the condition
on both variables $\delta$ and $B$, while 
the former does not impose any condition on the barrier value. 
Nevertheless, since we are interested in computing the
first-crossing distribution in the large scale limit, 
the barrier trajectories converge toward a unique constant value, 
$B_0=\delta_c$ for $S'\rightarrow
0$ and thus we recover $\mathcal{F}(S|\delta_0,0)$. 
Nevertheless, some care is needed when computing the non-Markovian corrections.

Following \cite{Ma2011} the path-integral defition of
the conditional first-crossing distribution is
\begin{equation}
\mathcal{F}(S_n|Y_m,S_m)=-\int_0^\infty dY_n \frac{\partial P(Y_n,S_n|Y_m,S_m)}{\partial S_n},\label{condf}
\end{equation}
where
\begin{equation}\label{pcond}
\begin{split}
&P(Y_n,S_n|Y_m,S_m)\equiv\\
&\equiv\frac{\int_0^\infty dY_1...d\hat{Y}_m...dY_{n-1}
  W(Y_0=\delta_c,...\hat{Y}_m=0,...,Y_n,S_n)}{\int_0^\infty dY_1...dY_{m-1}
  W(Y_0=\delta_c,...,Y_m,S_m)},
\end{split}
\end{equation}
with the probability density developped to first order in $\kappa$
reads as
\begin{equation}
W(...)=W_{0}(...)+\frac{1}{2}\sum_{i,j=1}^n\Delta_{ij}\partial_i\partial_j W_0(...),\label{wtot}
\end{equation}
and $W_0(...)$ is the probability density distribution of the discrete
Markovian random walks. Notice that the integral in the denominator of
Eq.~(\ref{pcond}) provides the correct normalization factor to the
conditional first-crossing distribution. The numerator in
Eq.~(\ref{pcond}) can be computed by splitting the sum
in Eq.~(\ref{wtot}) and computing each term
individually. However, as shown in \cite{Ma2011} only few of these 
terms actually contribute to $\mathcal{F}(S_n|Y_m,S_m)$ and we have verified
this to be the case also for the diffusive barrier model with linearly
drifting average. In particular we have
\begin{equation}\label{pcondtot}
\begin{split}
P&(Y_n,S_n|Y_m,S_m)=\Pi_{0}(Y_m,Y_n,S_n-S_m)+\\
&+\Pi^a_{1}(Y_m,Y_n,S_n-S_m)+\frac{N_{1}^b(Y_m,Y_n,S_m,S_n)}{\Pi_0(Y_0=\delta_c,Y_m,S_m)}
\end{split}
\end{equation}
where $\Pi_{0}(Y_m,Y_n,S_n-S_m)$ and $\Pi_0(Y_0=\delta_c,Y_m,S_m)$ are
given by the Markovian probability distribution Eq.~(\ref{pimarkov}). 
The other two terms in Eq.~(\ref{pcondtot})
contains the non-Markovian corrections to first order in $\kappa$, these
read as
\begin{widetext}
\begin{equation}\label{pi1a}
\begin{split}
\Pi^a_{1}(Y_m,Y_n,S_n-S_m)&=\frac{1}{2}\sum_{i,j=m+1}^{n-1}\Delta_{ij}\int_0^\infty
  dY_{m+1}...\int_0^\infty dY_{n-1}\partial_i\partial_j
  W_0(Y_m,...,Y_n,S_n-S_m)=\\
&=\sum_{i,j=m+1}^{n-1}\,\Delta_{ij}\Pi_{0}^\epsilon(Y_m,0,S_i-S_m)\Pi_{0}^\epsilon(0,0,S_j-S_i)\Pi_{0}^\epsilon(0,Y_n,S_n-S_j),
\end{split}
\end{equation}
and
\begin{equation}\label{N1b}
\begin{split}
N_{1}^b&(Y_m,Y_n,S_n,S_m)=\sum_{j=m+1}^{n-1}\Delta_{jm}\int_0^\infty dY_{1}...d\hat{Y}_j...\int_0^\infty
  dY_{n-1}\partial_m W_0(\delta_c,...,Y_m,S_m) W_0(Y_m,...,Y_j,S_j-S_m)
  \times\\&\times W_0(Y_j,...,Y_n,S_n-S_j)=\partial_m
  \Pi_0(Y_0=\delta_c,Y_m,S_m) \sum_{j=m+1}^{n-1}\Delta_{jm}\Pi_{0}^\epsilon(Y_m,0,S_j-S_m)\Pi_{0}^\epsilon(0,Y_n,S_n-S_j).
\end{split}
\end{equation}
\end{widetext}
As in the case of the non-Markovian corrections to the mass function,
these terms can be computed in the continous limit with the sum
sustituted with an integral over the variance and
the integrands given by Eq.~(\ref{atb}), (\ref{fromb}) and
(\ref{fromtob}) respectively.

The first term in Eq.~(\ref{pcondtot}) gives the Markovian conditional
first-crossing distribution. 
In the limit $S_m\rightarrow 0$ (i.e. $Y_m\rightarrow\delta_c-\delta_0$) we find
\begin{equation}
\mathcal{F}_0(S|\delta_0,0)=\frac{(\delta_c-\delta_0)}{S^{3/2}}\sqrt{\frac{a}{2\pi}}e^{-a\frac{(\delta_c-\delta_0+\beta
    S)^2}{2S}}
\end{equation}
Eq.~(\ref{pi1a}) is equivalent to the `memory-of-memory' term
Eq.~(\ref{memofmem}) In the limit $S_m\rightarrow 0$.
The double integral over the variance can be computed analytically by 
Taylor expanding in $\beta$. Here, we limit the computations to terms up
to first order in $\beta$, since its numerically calibrated value is
$\approx 10^{-2}$. We find
\begin{equation}
\begin{split}
\mathcal{F}_{1,\beta=0}^a&(S|\delta_0,0)=-\tilde{\kappa}\,\frac{(\delta_c-\delta_0)}{S^{3/2}}\sqrt{\frac{a}{2\pi}}\times\\
&\times\left[e^{-a\frac{(\delta_c-\delta_0)^2}{2S}}-\frac{1}{2}\Gamma\left(0,\frac{a(\delta_c-\delta_0)^2}{2S}\right)\right]
\end{split}
\end{equation}
and
\begin{equation}
\begin{split}
\mathcal{F}_{1,\beta^{(1)}}^a(S|\delta_0,0)&=-\beta\,a\,(\delta_c-\delta_0)\biggl[\mathcal{F}_{1,\beta=0}^a(S|\delta_0,0)+\\
&+\frac{\tilde{\kappa}}{2
  S}\textrm{Erfc}\left(\frac{\delta_c-\delta_0}{\sqrt{2
  S}}\sqrt{a}\right)\biggr]
\end{split}
\end{equation}

On the other hand, the computation of Eq.~(\ref{N1b}) requires some
care. This term contains the derivative of the Markovian solution,
$\Pi_0(\delta_c,Y_m,S_m)$, with respect to $Y_m$, hence
it explicitely depends on the variation of the probability distribution with respect
to the distance of the barrier $B_m$ to $\delta_m$. Nonetheless, we are interested in the
first-crossing distribution which is conditional in $\delta_m$ only,
and for which the result of the integration of Eq.~(\ref{N1b}) in the limit
$S_m\rightarrow 0$ should depend only on the variation with respect to
$\delta_m$. However, after computing this term
we find that for $\beta=D_B=0$ we do not recover the
non-Markovian correction of spherical collapse model. 
This is a direct consequence of the fact that we are working with $Y$
rather than $\delta$. The correct way to proceed is to
first marginalize over $B_m$ and then take the limit $S_m\rightarrow
0$. However, this is a very cumbersome computation, instead we have found that
the inconsistency can be cured by simply taking the derivative in Eq.~(\ref{N1b}) with
respect to $\delta_m$ only. In such a case we find
\begin{equation}
\begin{split}
\mathcal{F}_{1,\beta=0}^b(S|\delta_0,0)&=-\tilde{\kappa}\frac{\delta_0}{S^{3/2}}\sqrt{\frac{a}{2\pi}}\times\\
&\times e^{-a\frac{(\delta_c-\delta_0)^2}{2S}}\biggl\{1-a\frac{(\delta_c-\delta_0)^2}{S}\biggr\},
\end{split}
\end{equation}
which for $D_B=0$ coincides with the result of \cite{Ma2011} and 
\begin{equation}
\begin{split}
\mathcal{F}_{1,\beta^{(1)}}^b(S|\delta_0,0)&=-\beta\,a\,
\tilde{\kappa}\frac{\delta_0(\delta_c-\delta_0)}{S^{3/2}}\,2\sqrt{\frac{a}{2\pi}}\times\\
&\times e^{-a\frac{(\delta_c-\delta_0)^2}{2
S}}\biggl\{1-a\frac{(\delta_c-\delta_0)^2}{S}\biggr\},
\end{split}
\end{equation}

\begin{figure}[t]
\includegraphics[scale=0.42]{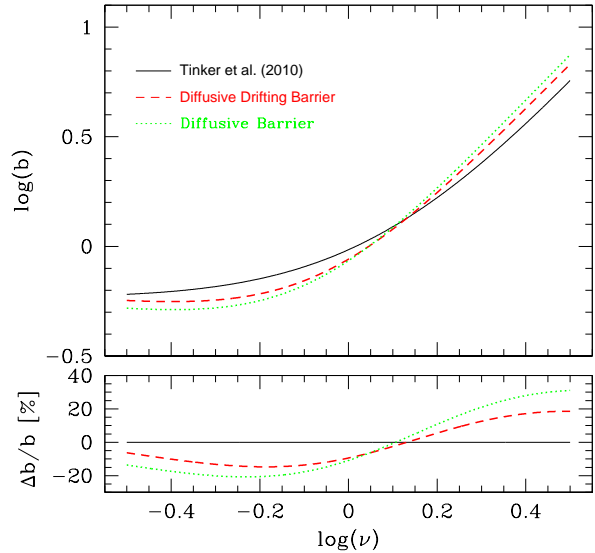}
\caption{Top panel: Halo bias for the diffusive drifting barrier model
with parameters best fitting mass function data from \cite{Tinker2008}
at $z=0$ (red short dash line) against the best fit formula to the
halo bias from the same numerical simulation sets inferred in
\cite{Tinker2010} (black solid line). Lower panel: relative difference
with respect to the halo fit formula from \cite{Tinker2010}.}
\label{fig9}
\end{figure}

Then, summing all these terms and evaluating Eq.~(\ref{bias}) we finally
obtain
\begin{equation}
\begin{split} 
&b_h(\nu)=1+\frac{1}{\delta_c}\frac{a\,\nu^2-1+\frac{\tilde{\kappa}}{2}\left[2-e^{\frac{a\nu^2}{2}}\Gamma\left(0,\frac{a\nu^2}{2}\right)\right]}{1-\tilde{\kappa}+\frac{\tilde{\kappa}}{2}e^{\frac{a\nu^2}{2}}\Gamma\left(0,\frac{a\nu^2}{2}\right)}+\\
&+\frac{\beta\,a}{\biggl[1-\tilde{\kappa}+\frac{\tilde{\kappa}}{2}e^{\frac{a\nu^2}{2}}\Gamma\left(0,\frac{a\nu^2}{2}\right)\biggr]^2}\times\biggl\{1-\tilde{\kappa}\biggl[a\nu^2+2\\
&-e^{\frac{a\nu^2}{2}}\Gamma\left(0,\frac{a\nu^2}{2}\right)-e^{\frac{a\nu^2}{2}}\sqrt{\frac{a\pi}{2}}\textrm{Erfc}\left(\frac{\nu\sqrt{a}}{2}\right)\biggr]\biggr\},
\end{split}
\end{equation}
where $\nu=\delta_c/\sigma$. In Fig.~\ref{fig9} we plot the halo bias for $\beta$ and
$D_B$ best fittting the mass function data at $z=0$ (red short dash
line) against the best fit formula to the bias measurements
inferred in \cite{Tinker2010} (black solid line). As we can see the
difference is $\lesssim 20\%$ over the mass range probed by the
simulations. This is consistent with the findings of Ma et
al. \cite{Ma2011}. As argued in \cite{Tinker2010} the discrepancy with
respect to the halo bias data from N-body simulations is
related to the peak-background split approximation itself. I
Hence, the improvement on mass function calculation 
does not give any further insight on the linear halo bias.

\section{Conclusions}\label{conclu}
The Excursion Set formalism provides a powerful mathematical framework
which allows us to perform a theoretical computation of the halo mass
function from a limited set of initial assumptions. These must
involve the statistics of the linear density fluctuation field as well
as a stochastic barrier model of the halo collapse conditions.
In addition, such calculation needs to be implemented with a
path-integral evaluation of the corrections due to the filtering of
the linear density field associated with a realistic mass definition.
Such an approach allows for a consistent model comparison
with N-body simulation data.

Here, we have derived an analytical expression for the mass function
and linear bias in the case of a diffusive barrier model with linearly
drifting average. This model well approximates the main features of
the ellipsoidal collapse. We have found a remarkable agreement with
N-body simulation data with differences $\approx 5 \%$ over a large range of masses. 
Such an agreement is due to the competing effects of the barrier average drift 
at small masses and of the diffusion in the high-mass end. 
This has important phenomenological implications especially in 
the study of primordial non-Gaussianity. In fact, several studies have
estimated the halo mass function in the case of non-Gaussian initial conditions
assuming the spherical collapse models (see e.g. \cite{MatarreseVerdeJimenez2000,Loverde2008,MaggioreRiotto2010c}). 
However the comparison with non-Gaussian N-body simulations has shown
large deviations in the low mass range compared to large masses (see e.g. Fig. 1 in
\cite{GiannantonioPorciani2010}). In the light of our results, it is
plausible that such discrepancies may be attributed to the
non-spherical collapse of halos. The inclusion of a simple diffusive
barrier model with linearly drifting average for a non-Gaussian linear
density field could resolve or alleviate the problem.

Our results suggest a number of directions which warrent further investigation.
Firstly, it will be insightful to derive the statistical properties of the
fuzzy barrier for a given ellipsoidal model as function of the variance of the linear density
field. This will provide theoretical predictions for $\beta$ and
$D_B$ which can be confronted with numerically calibrated values
for different redshifts and cosmologies, and it will allow us to
accurately testing the modeling of the halo collapse condtions.
On the other hand, in the upcoming years several observational campaigns will
probe the halo abundance through galaxy cluster surveys.
The mass function derived here can be used to perform a data
analysis of the barrier model parameters, thus providing
information on the collapse of DM halos which has been previously
unforseen.
 
\begin{acknowledgments}
We are especially thankful to J. Tinker for kindly providing us with
the mass function data. It is a pleasure to thank J.-M. Alimi, L. Amendola, M. Maggiore,
Y. Rasera, T. Riotto and R. Sheth for useful discussions. I. Achitouv is supported by a scholarship of the 
`Minist\`ere de l'Education Nationale, de la Recherche et de la Technologie' (MENRT). 

\end{acknowledgments}

\appendix

\section{Chapman-Kolmogorov Equation}\label{app0}
The probability density distribution of the 
discrete Markovian random walk is given by Eq.~(\ref{wdensity}) which by
expliciting the variance dependence of the connected $1$ and $2$-point
correlators reads as
\begin{equation}
W_0(Y_0,..,Y_n,S_n)=\int\mathcal{D}\lambda\,e^{i\sum_i\lambda_i
  (Y_i-\bar{B}_i)-\frac{\epsilon}{2}\sum_{ij}\lambda_i\lambda_j
  \tilde{A}_{ij}},\label{densprob1}
\end{equation}
where $\tilde{A}_{ij}=(1+D_B)\textrm{min}(i,j)$. Diagonalizing the
quadratic form in Eq.~(\ref{densprob1}) and solving
the resulting Gaussian integral we obtain
\begin{equation}
W_0(Y_0,..,Y_n,S_n)=\frac{e^{-\frac{1}{2\epsilon(1+D_B)}\sum_{ij}(Y_i-\bar{B}_i)A^{-1}_{ij}(Y_j-\bar{B}_j)}}{[2\pi\epsilon(1+D_B)]^\frac{n}{2}},\label{densprob2}
\end{equation}
where $A_{ij}=\textrm{min}(i,j)$. One can show through induction that $(A^{-1})_{ij}=2$ for
$i=1,...,n-1$, $(A^{-1})_{nn}=1$ and
$(A^{-1})_{i\,i+1}=(A^{-1})_{i+1\,i}=-1$ for $i=1,...,n-1$, while all
other elements vanish. Similarly one finds that
$\textrm{det}\,A=1$. Thus, we can write Eq.~(\ref{densprob2}) as 
\begin{equation}
\begin{split}
W_0(Y_0,&..,Y_n,S_n)=\frac{e^{-\frac{1}{2\epsilon(1+D_B)}\sum_{i=1}^{n-1}\left[(Y_{i+1}-Y_{i})-(\bar{B}_{i+1}-\bar{B}_{i})\right]^2}}{[2\pi\epsilon(1+D_B)]^\frac{n}{2}}\\\\
&=\psi_\epsilon(\Delta Y)\,W_0(Y_0,..,Y_{n-1},S_{n-1}),\label{wdecomp}\\
\end{split}
\end{equation}
with $\Delta{Y}=Y_n-Y_{n-1}$ and
\begin{equation}
\psi_\epsilon(\Delta Y)=\frac{1}{\sqrt{2\pi\epsilon(1+Db)}}e^{-\frac{(\Delta{Y}-\beta\,\epsilon)^2}{2\epsilon(1+D_B)}}.
\end{equation}

An important consequence of the above relations is that the Markovian
probability density satisfies the relation:
\begin{equation}
\begin{split}
W_0(Y_0,...,\hat{Y}_i=0,&...,Y_n,S_n)=W_0(Y_0,...,Y_{i-1},0,S_i)\times\\
&\times W_0(0,Y_{i+1},...,Yn,S_n-S_i)\label{derivw}
\end{split}
\end{equation}

Finally, from Eq.~(\ref{pepsilon}) we obtain the relation  
\begin{equation}
\Pi_0^\epsilon(Y_0,Y_n,S_n)=\int_0^\infty dY_{n-1} \,\psi_\epsilon(\Delta{Y})\,\Pi_0^\epsilon(Y_0,Y_{n-1},S_{n-1}),\label{chapman}
\end{equation}
which is the Chapman-Kolmogorov equation for a Gaussian random walk
with linearly drifting average.

\section{Calculation of $\gamma$}\label{app1}
We can compute the factor $\gamma$ defined by Eq.~(\ref{gamma}) simply
using the properties of the Markovian solution to the Fokker-Planck
equation Eq.~(\ref{FP}). Without loss of generality we can consider
the case where the absorbing boundary is at $Y=Y_c<0$ rather than
$Y=0$. In such a case the solution to Eq.~(\ref{FP})
with initial condition at $Y=Y_0>Y_c$ reads as
\begin{equation}\tilde{\Pi}=\frac{e^{a\,\beta(Y-Y_0-\beta\frac{S}{2})}\sqrt{a}}{\sqrt{2\pi
      S}}\left[e^{-\frac{a(Y-Y_0)^2}{2S}}-e^{-\frac{a(2 Y_c-Y-Y_0)^2}{2S}}\right],
\end{equation}
where $a=1/(1+D_B)$. Taking the derivative with respect to $Y_c$ we have 
\begin{equation}
\frac{\partial\tilde{\Pi}}{\partial Y_c}=a\sqrt{\frac{2\,a}{\pi}}\frac{2\,Y_c-Y_0-Y}{S^{3/2}}e^{a\,\beta(Y-Y_0-\beta\frac{S}{2})}e^{-\frac{a\,(2\,Y_c-Y_0-Y)^2}{2S}}.\label{dpidyc}
\end{equation}
On the other hand using the path-integral formulation we have 
\begin{equation}
\tilde{\Pi}_\epsilon=\int_{Y_c}^\infty dY_1...\int_{Y_c}^\infty dY_{n-1} W(Y_0,...,Y_n,S_n),
\end{equation}
thus
\begin{equation}
\frac{\partial\Pi_\epsilon}{\partial Y_c}=-\sum_i \int_{Y_c}^\infty
dY_1...\int_{Y_c}^\infty dY_{n-1}W(Y_0,...,\hat{Y}_i,...,Y_n,S_n),
\end{equation}
then using Eq.~(\ref{derivw}) and taking the continous limit we obtain
\begin{equation}
\frac{\partial\tilde{\Pi}}{\partial Y_c}=-\lim_{\epsilon\rightarrow
  0}\frac{1}{\epsilon}\int_0^S dS_i \Pi_\epsilon(0,Y_c,S_i)\Pi_\epsilon(Y_c,Y,S-S_i),\label{eqgamma}
\end{equation}
where the left-hand side is given by Eq.~(\ref{dpidyc}), while 
$\tilde{\Pi}_\epsilon(0,Y_c,S_i)$ and $\tilde{\Pi}_\epsilon(Y_c,Y,S-S_i)$ are  
the finite-$\epsilon$ corrections to the Markovian solution near the
boundary. As shown in Section~\ref{pathint} these terms can be evaluated by
introducing the stretch variable $\eta\sqrt{2\epsilon/a}$ and
expanding the Markovian solution to lowest order in $\epsilon$, this gives
\begin{equation}
\tilde{\Pi}_\epsilon(Y_0,Y_c,S)=\sqrt{\epsilon}\frac{\gamma a\,(Y_c-Y_0)}{S^{3/2}}e^{-\frac{a(Y_0-Y_c+\beta S)^2}{2 S}},
\end{equation}
and
\begin{equation}
\tilde{\Pi}\epsilon(Y_c,Y,S)=\sqrt{\epsilon}\frac{\gamma
  a\,(Y_c-Y)}{S^{3/2}}e^{-\frac{a(Y-Y_c-\beta S)^2}{2 S}}.
\end{equation}
Substituting these solutions on the right-hand-side of
Eq.~(\ref{eqgamma}) we have
\begin{equation}\label{dpidycint}
\begin{split}
\frac{\partial\tilde{\Pi}}{\partial
  Y_c}&=-(a\,\gamma)^2(Y_c-Y_0)(Y_c-Y)e^{a\,\beta(Y-Y_0-\beta\frac{S}{2})}\times\\
&\times\int_0^{S_n}dS_i\frac{e^{-\frac{a(Y-Y_c)^2}{2(S-Si)}}e^{-\frac{a(Y_0-Y_c)^2}{2S_i}}
   }{S_i^{3/2}(S-S_i)^{3/2}},
\end{split}
\end{equation}
the integral in $dS_i$ can be computed analytically since
\begin{equation}
\int_0^S
dS_i\frac{e^{-\frac{A^2}{2S_i}-\frac{B^2}{2(S-S_i)}}}{S_i^{3/2}(S-S_i)^{3/2}}=\sqrt{2\pi}\frac{A+B}{A
B}\frac{1}{S^{3/2}}e^{-\frac{(A+B)^2}{2S}},
\end{equation}
then, substituting in Eq.~(\ref{dpidycint}) and comparing with
Eq.~(\ref{dpidyc}) after simplifications we obtain $\gamma=1/\sqrt{\pi}$.

\section{Non-Markovian Corrections}\label{app2}
Using Eq.~(\ref{sum}) the non-Markovian
correction to first order in $\kappa$, Eq.~(\ref{nmcepsilon}), has two
separate contributions
\begin{widetext}
\begin{equation}
\begin{split}
&\Pi^{\textrm{m}}_{\epsilon,1}(Y_0,Y_n,S_n)=\sum_i\Delta_{in}\partial_n\biggl[\int_0^\infty
  dY_{1}...\int_0^\infty dY_{n-1}\partial_i W_0(Y_0,..,Y_n,S_n)\biggr]\\
&=-\sum_i\Delta_{in}\partial_n\biggl[\int_0^\infty
  dY_{1}...d\hat{Y}_i...\int_0^\infty
  dY_{n-1}W_0(Y_0,..,\hat{Y}_i=0,S_i)\,W_0(\hat{Y}_i=0,..,Y_n,S_n-S_i)\biggr]\\
&=-\sum_i\Delta_{in}\partial_n\biggl[\Pi_0^\epsilon(Y_0,0,S_i)\,\Pi_0^\epsilon(0,Y_n,S_n-S_i)\biggr].
\end{split}
\end{equation}
\end{widetext}
where we have used Eq.~(\ref{derivw}), and
\begin{widetext} 
\begin{equation}\label{memmemapp}
\begin{split}
\Pi^{\textrm{m-m}}_{\epsilon,1}&(Y_0,Y_n,S_n)=\sum_{i<j}\Delta_{ij}\int_0^\infty
  dY_{1}...\int_0^\infty dY_{n-1}\partial_i\partial_j \biggl[W_0(Y_0,..,Y_n,S_n)\biggr]\\
&=-\sum_i\Delta_{in}\int_0^\infty
  dY_{1}...d\hat{Y}_j...\int_0^\infty
  dY_{n-1}\partial_i\biggl[{W_0(Y_0,..,\hat{Y}_j=0,..,Y_n,S_n)}\biggr]\\
&=\sum_{i<j}\Delta_{ij}\int_0^\infty
  dY_{1}...d\hat{Y}_i...d\hat{Y}_j...\int_0^\infty
  dY_{n-1}\biggl[{W_0(Y_0,..\hat{Y}_i=0,..,\hat{Y}_j=0,S_j)\,W_0(\hat{Y}_i=0,..,Y_n,S_n-S_j)}\biggr]\\
&=\sum_{i<j}\Delta_{ij}\biggl[\Pi_0^\epsilon(Y_0,0,S_i)\,\Pi_0^\epsilon(0,0,S_j-S_i)\,\Pi_0^\epsilon(0,Y_n,S_n-S_j)\biggr].
\end{split}
\end{equation}
\end{widetext}

\end{document}